# Digital Watermarking for Image Authentication Based on Combined DCT, DWT and SVD Transformation

Mohammad Ibrahim Khan[1], Md. Maklachur Rahman[2] and Md. Iqbal Hasan Sarker[3]

[1]Department of Computer Science and Engineering, Chittagong University of Engineering and Technology
Chittagong-4349, Bangladesh

[2]Department of Computer Science and Engineering, Chittagong University of Engineering and Technology
Chittagong-4349, Bangladesh

[3]Department of Computer Science and Engineering, Chittagong University of Engineering and Technology
Chittagong-4349 Bangladesh

**Abstract**
Digital content can frequently copied by unauthorized person and claim to his ownership. But we don't know who the actual owner of that content is. Digital Watermarking is an important issue to solve this kind of problem.
This paper presents a hybrid digital image watermarking based on Discrete Wavelet Transform (DWT), Discrete Cosine Transform (DCT) and Singular Value Decomposition (SVD) in a zigzag order. From DWT we choose the high band to embed the watermark that facilities to add more information, gives more invisibility and robustness against some attacks. Such as geometric attack. Zigzag method is applied to map DCT coefficients into four quadrants that represent low, mid and high bands. Finally, SVD is applied to each quadrant.

## 1. Introduction

Digital watermarking describes a method and technology that hide information. It is the process of embedding a piece of digital information into any multimedia data. In some watermarking schemes, a Watermarked image has a logo or some other information embedded into the image that may visible or invisible by people. The quality of the watermarking Scheme largely depends upon the choice of the watermark structure and insertion strategy. There are basically two approaches to embed a watermark: spatial domain and transform domain watermarking. The two main constraints involved in the problem of watermarking are those of maintaining the robustness of the watermark information while keeping visual perception of the original image intact. Apart from copy control and copyright protection; broadcast monitoring, fingerprinting, indexing, medical application, content authentication are other application areas of digital watermarking. For the purpose of designing and developing a new watermarking algorithm in those application areas, the most important properties are robustness and invisibility which are the main point of this study.

## 2. Overview of Transforms for Watermarking

2.1 Discrete Wavelet Transform (DWT)

Wavelet transform decomposes an image into a set of band limited components which can be reassembled to reconstruct the original image without error. For 2-D images, applying DWT corresponds to processing the image by 2-D filters in each dimension. The filters divide the input image into four non-overlapping multi-resolution sub bands, a lower resolution approximation image (LL1), horizontal (HL1), vertical (LH1) and diagonal (HH1) detail components. The process can be repeated to obtain multiple scale wavelet decomposition. The information of low frequency district is an image close to the original image. Most signal information of original image is in this frequency district. The frequency districts of LH, HL and HH respectively represents the level detail, the upright detail and the diagonal detail of the original image. According to the character of HVS, human eyes are sensitive to the change of smooth district of image, but not sensitive to the tiny change of edge, profile and streak. Embedding the watermark in the higher level sub bands increases the robustness of the watermark. However, the image visual fidelity may be lost, which can be measured by PSNR. With the DWT, the edges and texture can be easily identified in the high frequency band .Therefore it's hard to conscious that putting the watermarking signal into the big amplitude coefficient of high-frequency band of the image DWT transformed. Then it can carry more watermark signal and has good concealing effect.

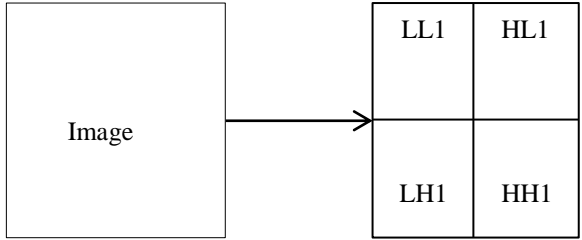

Fig. 1  Single level DWT

## 2.2 Discrete Cosine Transform (DCT)

The discrete cosine transform (DCT) is a way to transform a signal into elementary frequency components. Two dimensional DCT is used in image compression. In which the 2-D DCT of a given matrix gives the frequency coefficients in form of another matrix where vertical and horizontal dimensions are considered. DCT-based watermarking is based on two facts. The first fact is that much of the signal energy lies at low-frequencies sub-band which contains the most important visual parts of the image. The second fact is that high frequency components of the image are usually removed through compression and noise attacks. In below formulae for calculating DCT is given by eqn. 1 and inverse DCT is given by eqn. 2.

Equation of 2-D DCT:

$$F(u,v) = \sum_{i=0}^{N-1}\sum_{j=0}^{N-1} C(u)C(v)f(i,j)\cos\left[\frac{\pi(2i+1)u}{2N}\right] * \cos\left[\frac{\pi(2j+1)u}{2N}\right] \quad (1)$$

Equation of 2-D inverse DCT:

$$f(i,j) = \sum_{i=0}^{N-1}\sum_{j=0}^{N-1} C(u)C(v)F(u,v)\cos\left[\frac{\pi(2i+1)u}{2N}\right] * \cos\left[\frac{\pi(2j+1)u}{2N}\right] \quad (2)$$

Where,

$$C(u), C(v) = \begin{cases} \sqrt{\frac{1}{N}}, & u,v = 0 \\ \sqrt{\frac{2}{N}}, & u,v = 1 \text{ to } N-1 \end{cases}$$

## 2.3 Singular Value decomposition (SVD)

SVD is an effective numerical analysis tool used to analyze matrices. In SVD transformation, a matrix can be decomposed into three matrices that are of the same size as the original matrix. From the view point of linear algebra, an image is an array of nonnegative scalar entries that can be regarded as a matrix. Without loss of generality, if A is a square image, denoted as $A \in R^{n \times n}$, where R represents the real number domain, then SVD of A is defined as $A = USV^T$ where U and V are orthogonal matrices, and S is a diagonal matrix, as

$$S = \begin{bmatrix} s_1 & & \\ & \ddots & \\ & & s_n \end{bmatrix}$$

Here diagonal elements i.e. s's are singular values and satisfy $s_1 \geq s_2 \geq \ldots s_r \geq s_{r+1} \geq \ldots = s_n = 0$

SVD is an optimal matrix decomposition technique in a least square sense that it packs the maximum signal energy into as few coefficients as possible.

## 3. The Proposed Scheme

In our proposed watermarking scheme, we have devised a DWT, DCT and SVD based hybrid watermarking scheme by utilizing the salient features of DWT, DCT and the SVD. In the following subsection, we have described the watermark embedding and extraction process by using flowchart for watermark embedding and watermark extraction respectively that shows how watermark image is embedded with host image and how the embedded watermark is extracted from the attacked watermarked image.

## 3.1 Watermark Embedding Process

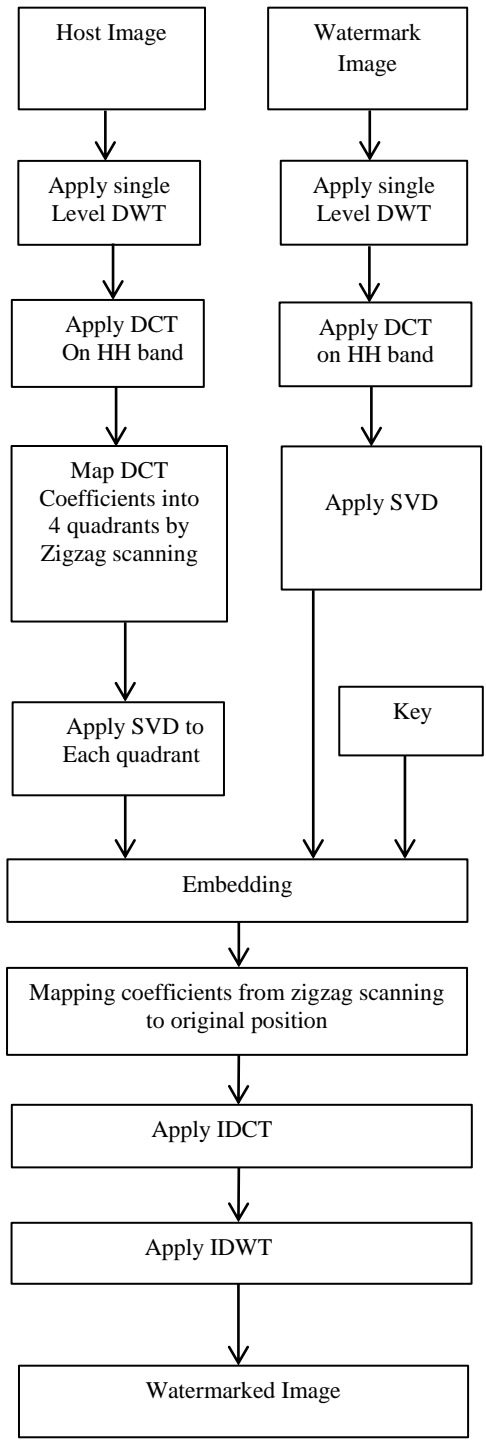

Fig. 2 Flowchart for Watermark Embedding

## 3.2 Watermark Extraction Process

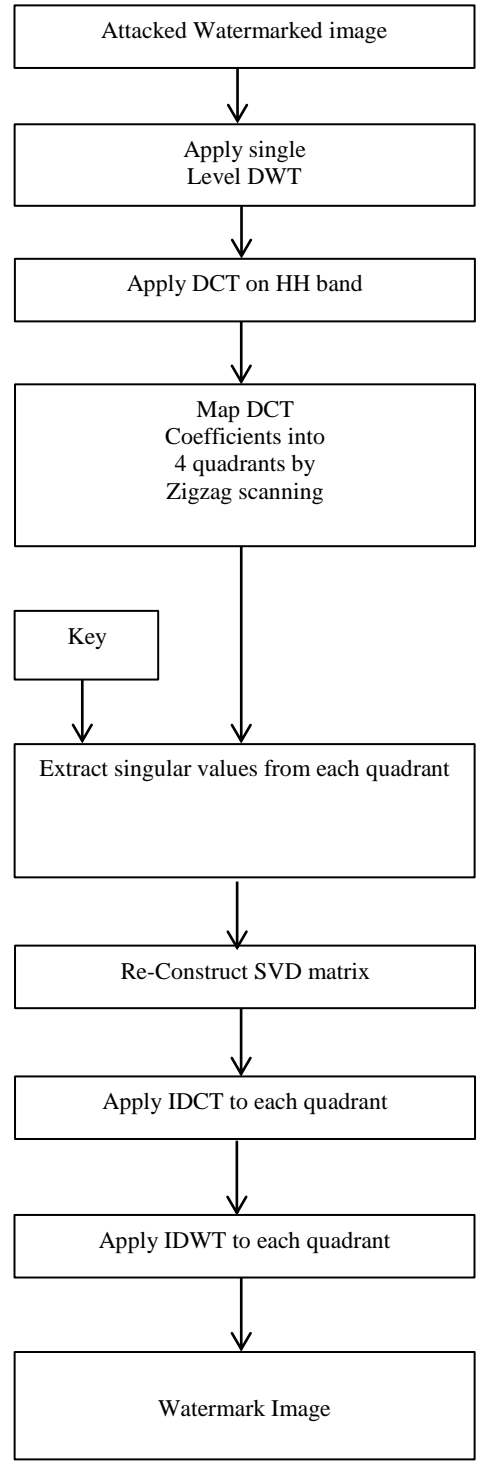

Fig.3 Flowchart for Watermark Extraction

### 3.3 Algorithm: Watermark Embedding

1. Input HI as Host image. Apply DWT to decompose it into four sub-bands LL , HL , LH and HH .

2. Select HH band and apply DCT to it and get DCT coefficient matrix H.

3. Map DCT coefficient matrix H into four quadrants $q_1$, $q_2$, $q_3$ and $q_4$ by using zigzag scanning.

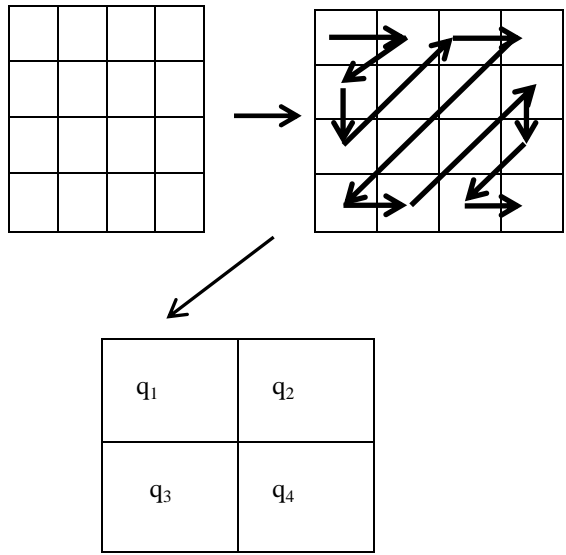

Fig. 4  Mapping DCT coefficients into four quadrants by zigzag scanning.

4. Apply SVD to each quadrant $q_1$, $q_2$, $q_3$ and $q_4$ to get S1, S2, S3 and S4.

5. Input wi as Watermark image. Apply DWT to decompose it into four sub-bands LL , HL , LH and HH .

6. Select HH band and apply DCT to it and get DCT coefficient matrix w.

7. Apply SVD on matrix w to get Sw.

8. Modify S1, S2, S3 and S4 by using equation
$S_{ii} = S_i + \alpha * S_w \ where, i = 1\ to\ 4$.

9. Mapping coefficients from zigzag scanning to original position matrix H*.

10.  Apply inverse DCT to H* to produce HH*.

11. Apply inverse DWT to LL, HL, LH and HH* to get watermarked image WI.

### 3.4 Algorithm: Watermark Extraction

1. Input WI as Watermarked image. Apply DWT to decompose it into four sub-bands LL , HL , LH and HH.

2. Select HH band and apply DCT to it and get DCT coefficient matrix W.

3. Map DCT coefficient matrix W into four quadrants $q_1$, $q_2$, $q_3$ and $q_4$ by using zigzag scanning.

4.  Modify S1, S2, S3 and S4 by using equation
$S_w = (S_{ii} - S_i)/\alpha \ where, i = 1\ to\ 4$.

5.  Re-construct SVD matrix for each quadrant $q_1$, $q_2$, $q_3$ and $q_4$.

6. Apply inverse DCT and inverse DWT to each quadrant.

## 4. Experimental results

In this proposed watermarking algorithm host image Bird of size 512x512 is used as shown in Fig.5. Watermark image taken cameraman of size 256x256 shown in Fig.6.

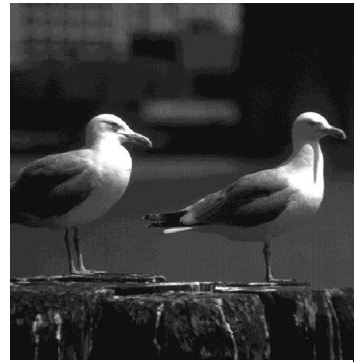 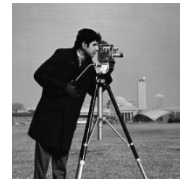

Fig. 5  Host image bird.          Fig. 6  watermark

The proposed watermarking algorithm is simulated using MATLAB 9. The proposed watermarking algorithm is tested for the various host and watermark images.  Here the results are given for Bird image only. To evaluate  the  performance  of the  proposed method, evaluation metrics used are

PSNR (Peak Signal to Noise Ratio) and NC (Normalized Correlation). PSNR is widely used to measure imperceptibility between the original image and watermarked image. PSNR is defined by the eqn. (3). The similarity between the original watermark and extracted watermark from the attacked image is evaluated by using NC given by the eqn. (4).

$$PSNR = 10 \log_{10}\left(\frac{255^2}{MSE}\right) \qquad (3)$$

Where,

$$MSE = \frac{1}{M \times N} \sum_{m=1}^{M} \sum_{n=1}^{N} [I(m,n) - I_w(m,n)]^2$$

$$NC = \frac{\sum_i \sum_j w(i,j)\, w'(i,j)}{\sum_i \sum_j |w(i,j)|^2} \qquad (4)$$

Fig. 7 shows different types of noisy attacked image and Fig.8 shows the extracted watermarked image from corresponding noisy attacked image. In which we will see how noises are effect on our watermarked images in our human eyes. And Table 1 shows normalized correlation (NC) values between the actual watermark and extracted watermark from attacked watermarked image. Table 2 shows the PSNR values of the host image and watermarked images with attack and also the normalized correlation (NC) values between the actual watermark and extracted watermark from attacked watermarked image for the proposed method and DWT, DCT and SVD method.

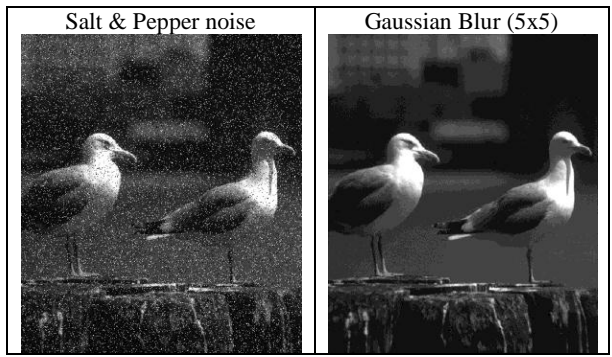

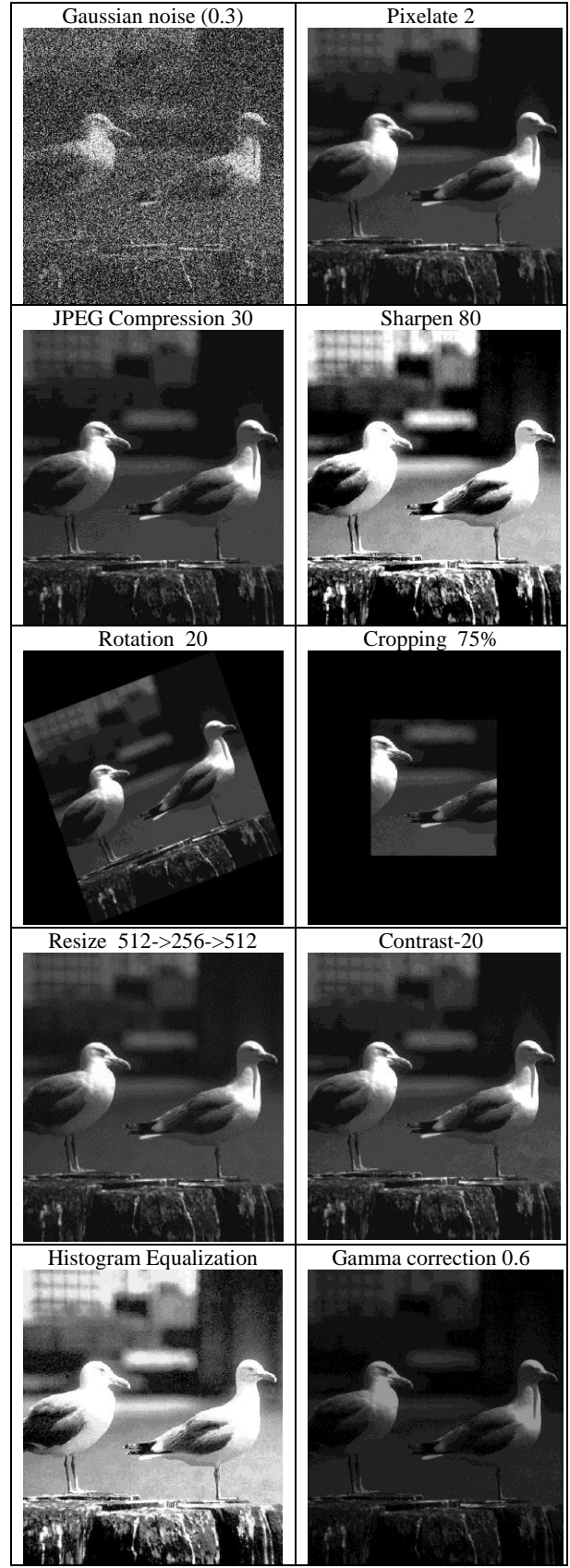

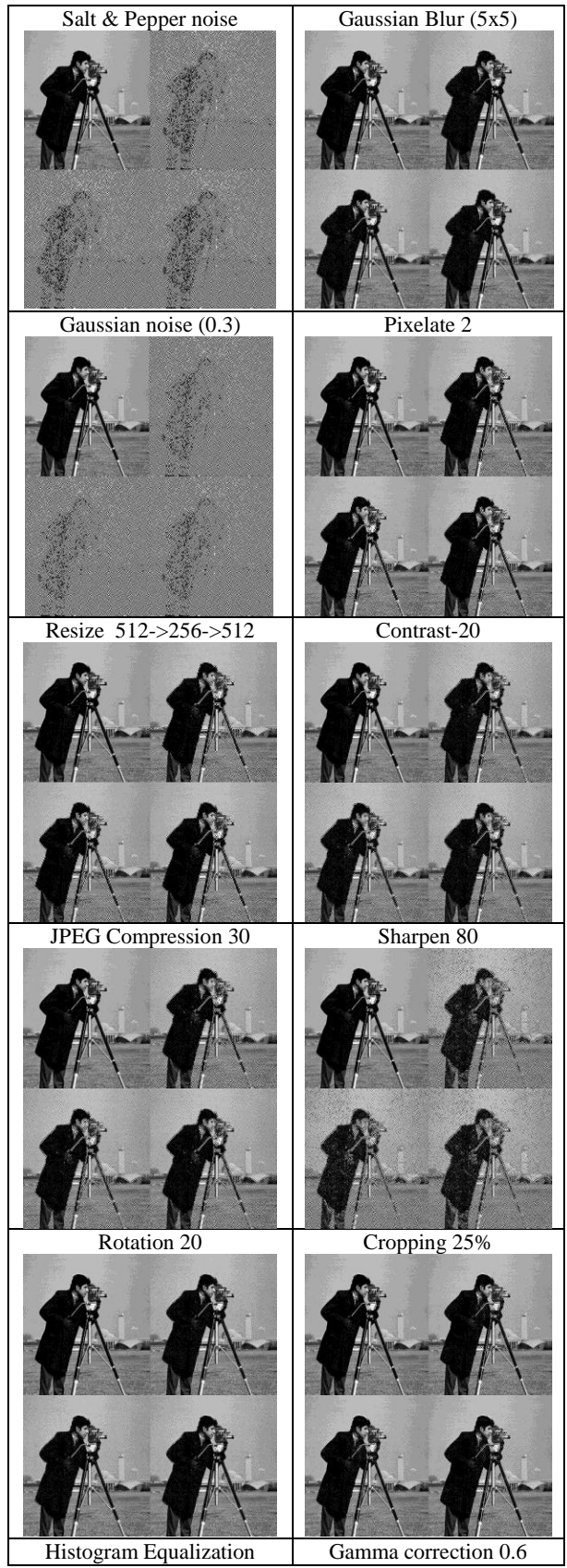

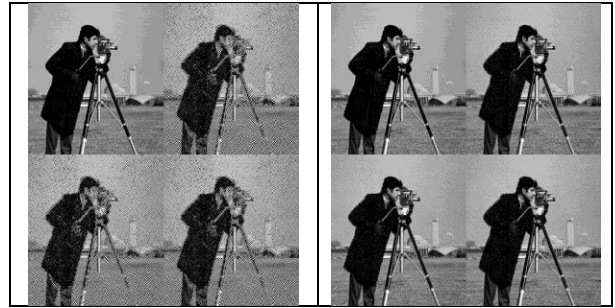

Fig. 7 Attacked watermarked and Extract watermark image

Table 1: Performance results in terms of NC values

| Attacks | Existing Method DCT-SVD (ref 1: Table 2) Max. NC | Proposed Method DWT-DCT-SVD Max. NC |
|---|---|---|
| Gaussian Blur (5x5) | 0.9917 | **0.9984** |
| JPEG Compression 30 | 0.9910 | **0.9983** |
| Sharpen 80 | 0.9811 | **0.9979** |
| Gaussian noise 0.3 | 0.9093 | **0.9762** |
| Pixelate 2 (Photoshop) | 0.9947 | **0.9986** |
| Rotation 20 | 0.9723 | **0.9985** |
| Cropping 25% area remaining | 0.9183 | **0.9985** |
| Resize 512->256->512 | 0.9937 | **0.9986** |
| Contrast-20 (Photoshop) | 0.9943 | **0.9983** |
| Histogram Equalization | 0.9874 | **0.9979** |
| Gamma correction 0.6 | **0.9995** | 0.9984 |
| Salt & Pepper noise | 0.9713 | **0.9894** |
| Poisson noise | ---- | 0.9981 |
| Speckle noise | ---- | 0.9981 |

Table 2: Performance results in terms of PSNR and NC values

| Attacks | Existing Method DWT(ref 7) | | Proposed Method DWT-DCT-SVD | |
|---|---|---|---|---|
| | PSNR | NC | PSNR | NC |
| JPEG 75 | 34.96 | 0.920 | **45.77** | **0.9983** |
| JPEG 50 | 33.11 | 0.840 | **35.93** | **0.9882** |
| JPEG 25 | 31.27 | 0.747 | **34.53** | **0.9983** |
| Blur 3x3 | 29.63 | 0.822 | **42.49** | **0.9984** |
| Gaussian noise[0 0.001] | 29.74 | 0.717 | **35.94** | **0.9982** |
| Resize 512->256->512 | **19.81** | 0.780 | 11.60 | **0.9986** |
| Histogram Equalization | **17.42** | 0.703 | 9.4008 | **0.9979** |
| Intensity Adj. ([0 0.8],[0 1]) | 18.87 | 0.883 | **25.54** | **0.9983** |
| Gamma Correction 1.5 | 17.90 | 0.908 | **18.88** | **0.9982** |
| Rotate 20 | 11.44 | 0.910 | **12.66** | **0.9985** |
| Cropping | 11.88 | 0.996 | **20.31** | **0.9983** |
| Pixelate 2 (Photoshop) | 30.13 | **1.000** | **32.47** | 0.9986 |
| Sharpen(Photoshop) | 31.21 | 0.839 | **33.95** | **0.9981** |
| Re-watermark | 38.51 | 0.905 | **43.04** | **0.9982** |
| Collusion | 45.35 | 0.867 | **49.31** | **0.9981** |

From Table 1 and 2, it is observed that the proposed DWT-DCT-SVD watermarking algorithm gives more PSNR and NC values than the DCT -SVD and DWT method. That gives more imperceptibility and robust against different kinds of noise.

## 5. Conclusion

In this paper, we described a combined DWT-DCT digital image watermarking algorithm where discrete wavelet transform(DWT), discrete cosine transform (DCT) Singular Value decomposition (SVD) and their cross combination have been applied successfully in many in digital image watermarking. The combination of the three transforms improved the watermarking performance considerably when compared to the DCT–SVD, DWT-Only watermarking approach. Further-more the proposed algorithm can be improved using DWT-DCT-SVD and further can be extended to color images and video processing.

## References

[1] P. S. Murty, K. S. Dileep and P. R. Kumar, "A Semi Blind Self Reference Image Watermarking in Discrete Cosine Transform using Singular Value Decomposition", International Journal of Computer Applications, vol. 62, issue 13, January 2013, pp. 29-36.
[2] S. K. Prajapati, A. Naik and A. Yadav, "Robust Digital Watermarking using DWT-DCT-SVD", International Journal of Engineering Research and Applications Vol. 2, Issue 3, May-Jun 2012, pp.991-997.
[3] Dr. M. A. Dorairangaswamy, "A Robust Blind Image Watermarking Scheme in Spatial Domain for Copyright Protection", International Journal of Engineering and Technology Vol. 1, No.3, August, 2009.
[4] A. Al-Haj, "Combined DWT-DCT Digital Image Watermarking", Journal of Computer Science 3 (9): 740-746, 2007.
[5] M. Calagna, H. Guo, L. V. Mancini and S. Jajodia, "A Robust Watermarking System Based on SVD Compression", Proceedings of ACM Symposium on Applied Computing (SAC2006), Dijon, France, pp. 1341-1347, 2006.
[6] F. Cayre, C. Fontaine and T. Furon, "Watermarking security: theory and practice", Signal Processing, IEEE Transactions on, vol. 53, no. 10, pp. 3976–3987, Oct. 2005.
[7] P. Taoaand and A. M. Eskicioglu, "A robust multiple watermarking scheme in the Discrete Wavelet Transform domain", Internet Multimedia Management Systems Proceedings of the SPIE, Volume 5601, pp. 133-144 (2004).
[8] F. Huang and ZH. Guan, "A hybrid SVD-DCT watermarking method based on LPSNR", Pattern Recognition Letters, vol.25, pp.1769 1775, 2004.
[9] Y. Wang, J. F. Doherty and R. E. Van Dyck, "A Wavelet-Based Watermarking Algorithm for Ownership Verification of Digital Images", IEEE Transactions on Image Processing, Volume 11, No. 2, February 2002, pp. 77-88.

[10] R. Liu and T. Tan, "A SVD-Based Watermarking Scheme for Protecting Rightful Ownership", IEEE Transactions on Multimedia, 4(1), March 2002, pp.121-128.
[11] S. D. Lin and C. F. Chen, "A Robust DCT-Based Watermarking for Copyright Protection",IEEE Transactions on Consumer Electronics, Volume 46, No. 3, August 2000, pp. 415-421.
[12] N. Nikolaidis and I.Pitas,"Robust image watermarking in the spatial domain", Signal Processing Vo1.66, pp.385-403, 1998.
[13] S. Mukherjee and A. K. Pal, "A DCT-SVD based Robust Watermarking Scheme for Grayscale Image", International Conference on Advances in Computing, Communications and Informatics (ICACCI-2012).
[14] N. H. Divecha, N. N. Jani, "Image Watermarking Algorithm using Dct, Dwt and Svd", IJCA Proceedings on National Conference on Innovative Paradigms in Engineering and Technology (NCIPET 2012), ncipet - Number 10.
[15] Z. Zhou, B. Tang and X. Liu, "A Block SVD Based Image Watermarking Method", Proceedings of the 6th World Congress on Intelligent Control and Automation, June 21 - 23, 2006, Dalian, China.
[16] E. Ganic and A. M. Eskicioglu, "Robust DWT-SVD Domain Image Watermarking: Embedding Data in All Frequencies", 13thEuropean Signal Processing Conference (EUSIPCO 2005), Antalya, Turkey, September 4-8, 2005.
[17] A. Sverdlov, S. Dexter and A. M. Eskicioglu, "Robust DCT-SVD Domain Image Watermarking for Copyright Protection: Embedding Data in All Frequencies", submitted to Multimedia Computing and Networking 2005 Conference, San Jose, CA, January 16-20, 2005.

**Dr. Mohammad Ibrahim Khan** received the B.S. degree in Electrical and Electronic Engineering from Bangladesh University of Engineering and Technology (BUET), Dhaka, Bangladesh in 1999. He received M.S. degree in Computer Science and Engineering from the same University in 2002. He received his Ph.D. degree in Computer Science and Engineering from Jahangirnagar University in 2010. Since 1999, he has been serving as a faculty member in the Department of Computer Science and Engineering at Chittagong University of Engineering & Technology (CUET), Chittagong, Bangladesh. His research interest includes Digital Image Processing, Graph Theory, Cryptography, Digital Watermarking, Multimedia Systems, and Digital Signal Processing.

**Md. Maklachur Rahman** will receive B.Sc. degree in Computer Science and Engineering (CSE) from Chittagong University of Engineering & Technology (CUET), Chittagong, Bangladesh in July, 2013. Currently he is a final year student Dept. of CSE. His research interest includes Digital Image Processing, Multimedia Security, Artificial Intelligence, Digital Watermarking and Software Engineering.

**Md. Iqbal Hasan Sarker** received the B.Sc. degree in Computer Science & Engineering from Chittagong University of Engineering & Technology (CUET), Chittagong, Bangladesh in 2009. Currently, he is pursuing M.Sc. in Computer Science & Engineering in the same University. Since 2010, he has been serving as a faculty member in the same department and university. His research interest includes Digital Watermarking, Computer Vision, Digital Image Processing, Cryptography and Data mining.